\global\def\draftcontrol{0}
   \def\versionno{ Boson Stars in AdS  }
\input{alex.tex}
\documentclass[12pt,letterpaper]{article}

\usepackage{amsmath,amssymb,array,calc,epsfig}
\usepackage{graphicx}
\usepackage{psfrag,verbatim,bm}
\usepackage{hyperref}
\usepackage{amssymb,graphicx}
\usepackage[usenames]{color}

\begin{document}

\ifnum\draftcontrol=1
\tolerance=1000
\fi

\renewcommand\baselinestretch{1.25}
\setlength{\paperheight}{11in}
\setlength{\paperwidth}{8.5in}
\setlength{\textwidth}{\paperwidth-2.4in}     \hoffset= -.3in   
\setlength{\textheight}{\paperheight-2.4in}   \topmargin= -.6in 

\renewcommand\section{\@startsection {section}{1}{\z@}%
                                   {-3.5ex \@plus -1ex \@minus -.2ex}%
                                   {2.3ex \@plus.2ex}%
                                   {\normalfont\large\bfseries}}
\renewcommand\subsection{\@startsection{subsection}{2}{\z@}%
                                   {-3.25ex\@plus -1ex \@minus -.2ex}%
                                   {1.5ex \@plus .2ex}%
                                   {\normalfont\normalsize\bfseries}}
\renewcommand\subsubsection{\@startsection{subsubsection}{3}{\z@}%
                                   {-3.25ex\@plus -1ex \@minus -.2ex}%
                                   {1.5ex \@plus .2ex}%
                                   {\normalfont\normalsize\it}}
\renewcommand\paragraph{\@startsection{paragraph}{4}{\z@}%
                                   {-3.25ex\@plus -1ex \@minus -.2ex}%
                                   {1.5ex \@plus .2ex}%
                                   {\normalfont\normalsize\bf}}

\numberwithin{equation}{section}



\def\ie{{\it i.e.}}
\def\eg{{\it e.g.}}

\def\revise#1       {\raisebox{-0em}{\rule{3pt}{1em}}%
                     \marginpar{\raisebox{.5em}{\vrule width3pt\
                     \vrule width0pt height 0pt depth0.5em
                     \hbox to 0cm{\hspace{0cm}{%
                     \parbox[t]{4em}{\raggedright\footnotesize{#1}}}\hss}}}}

\newcommand\fnxt[1] {\raisebox{.12em}{\rule{.35em}{.35em}}\mbox{\hspace{0.6em}}#1}
\newcommand\nxt[1]  {\\\fnxt#1}

\def\cala         {{\cal A}}
\def\calA         {{\mathfrak A}}
\def\calAbar      {{\underline \calA}}
\def\calb         {{\cal B}}
\def\calc         {{\cal C}}
\def\cald         {{\cal D}}
\def\cale         {{\cal E}}
\def\calf         {{\cal F}}
\def\calg         {{\cal G}}
\def\calG         {{\mathfrak G}}
\def\calh         {{\cal H}}
\def\cali         {{\cal I}}
\def\calj         {{\cal J}}
\def\calk         {{\cal K}}
\def\call         {{\cal L}}
\def\calm         {{\cal M}}
\def\caln         {{\cal N}}
\def\calo         {{\cal O}}
\def\calp         {{\cal P}}
\def\calq         {{\cal Q}}
\def\calr         {{\cal R}}
\def\cals         {{\cal S}}
\def\calt         {{\cal T}}
\def\calu         {{\cal U}}
\def\calv         {{\cal V}}
\def\calw         {{\cal W}}
\def\calz         {{\cal Z}}

\def\complex      {{\mathbb C}}
\def\naturals     {{\mathbb N}}
\def\projective   {{\mathbb P}}
\def\rationals    {{\mathbb Q}}
\def\reals        {{\mathbb R}}
\def\zet          {{\mathbb Z}}

\def\del          {\partial}
\def\delbar       {\bar\partial}
\def\ee           {{\rm e}}
\def\ii           {{\rm i}}
\def\chain        {{\circ}}
\def\tr           {\mathop{\rm Tr}}
\def\Re           {{\rm Re\hskip0.1em}}
\def\Im           {{\rm Im\hskip0.1em}}
\def\id           {{\it id}}

\def\de#1#2{{\rm d}^{#1}\!#2\,}
\def\De#1{{\cald}#1\,}

\def\half{{\frac12}}
\newcommand\topa[2]{\genfrac{}{}{0pt}{2}{\scriptstyle #1}{\scriptstyle #2}}
\def\undertilde#1{{\vphantom#1\smash{\underset{\widetilde{\hphantom{\displaystyle#1}}}{#1}}}}
\def\prodprime{\mathop{{\prod}'}}
\def\gsq#1#2{%
    {\scriptstyle #1}\square\limits_{\scriptstyle #2}{\,}} 
\def\sqr#1#2{{\vcenter{\vbox{\hrule height.#2pt
 \hbox{\vrule width.#2pt height#1pt \kern#1pt
 \vrule width.#2pt}\hrule height.#2pt}}}}
\def\square{%
  \mathop{\mathchoice{\sqr{12}{15}}{\sqr{9}{12}}{\sqr{6.3}{9}}{\sqr{4.5}{9}}}}

\newcommand{\fft}[2]{{\frac{#1}{#2}}}
\newcommand{\ft}[2]{{\textstyle{\frac{#1}{#2}}}}
\def\jsquare{\mathop{\mathchoice{\sqr{8}{32}}{\sqr{8}{32}}
{\sqr{6.3}{9}}{\sqr{4.5}{9}}}}

\newcommand{\wn}{\mathfrak{w}}


\def\a{\alpha}
\def\b{\beta}
\def\l{\lambda}
\def\w{\omega}
\def\dd{\delta}
\def\r{\rho}
\def\c{\chi}
\newcommand{\qq}{\mathfrak{q}}
\newcommand{\ww}{\mathfrak{w}}
\def\p{\phi}
\def\k{\kappa}
\def\hr{\hat{r}}
\def\om{\Omega}
\def\e{\epsilon}
\def\bm{\bar{\mu}}
\def\tc{\tilde{\calc}}
\def\tz{\tilde{z}}
\def\hz{\hat{z}}
\def\t{\tau}

\catcode`\@=12


\title{\bf Boson Stars in AdS}
\pubnum
{UWO-TH-13/5
}

\date\today

\author{
Alex Buchel,$^{1,2}$, Steven L. Liebling$^{1,3}$ and Luis Lehner$^{1}$\\[0.4cm]
\it $^1$\,Perimeter Institute for Theoretical Physics\\
\it Waterloo, Ontario N2J 2W9, Canada\\[0.2cm]
\it $^2$\,Department of Applied Mathematics\\
\it University of Western Ontario\\
\it London, Ontario N6A 5B7, Canada\\[0.2cm]
\it $^3$\,Department of Physics, Long Island University\\
\it Brookville, NY 11548, U.S.A.\\[0.2cm]
}

\Abstract{
We construct boson stars in global Anti de Sitter~(AdS) space
and study their stability. Linear perturbation results suggest that the ground
state along with the first three excited state boson stars are stable. We
evolve some of these solutions and study their nonlinear stability in light
of recent work~\cite{Bizon:2011gg} arguing that a weakly turbulent
instability drives scalar perturbations of AdS to black hole formation. 
However evolutions suggest that boson stars are nonlinearly stable and immune to 
the instability for sufficiently small perturbation. Furthermore, these studies find other families of initial data which similarly avoid the instability for sufficiently weak parameters. Heuristically, we argue that initial data families
with widely distributed mass-energy distort the spacetime sufficiently to
oppose the coherent amplification favored by the instability. 
From the dual CFT perspective our findings suggest 
that there exist families of rather generic initial conditions 
in strongly coupled CFT (with large number of degrees of freedom) 
that do not thermalize in the infinite future.
}

\makepapertitle

\body

\version\versionno

\section{Introduction}
Understanding the gravitational behavior of spacetimes which asymptotically
behave as anti deSitter (AdS) has, since holography~\cite{Aharony:1999ti}, 
attracted significant interest. The AdS/CFT correspondence conjecture implies
 that such understanding is critically important for a plethora of phenomena described
by field theories. Remarkably however, relatively little is known about
dynamical scenarios, especially in comparison to spacetimes which are asymptotically
flat~(AF) or asymptotically deSitter~(dS). An important reason behind this difference is that
the boundary of AdS
is in causal contact with the interior of the spacetime, in stark contrast to the boundaries
of AF and asymptotically dS spacetimes. That the boundary affects the interior prevents a
straightforward extension of
standard singularity theorems to  asymptotically AdS~(aAdS) spacetimes~\cite{Ishibashi:2012xk}.

Significant advances have recently been achieved via different approaches, including
strictly analytic~\cite{Holzegel:2011qj,Holzegel:2011rk,Holzegel:2012wt,Holzegel:2013kna},
perturbative~\cite{Dias:2012tq,Dias:2011ss} and 
numerical~\cite{Bizon:2011gg,Chesler:2010bi,Garfinkle:2011hm,Bantilan:2012vu,Buchel:2012uh,Chernicoff:2012gu,Maliborski:2013jca} 
(to name a few representative) efforts. 
Particularly intriguing is the evidence first presented
in~\cite{Bizon:2011gg} that pure AdS is unstable to scalar collapse to black hole~(BH), regardless
of how small a scalar perturbation is considered. 
The effect of the AdS boundary allows for the reflection of scalar pulses. Hence, a weak scalar pulse
bounces off the boundary at infinity, returning to concentrate again at the origin. The weakly turbulent
instability of~\cite{Bizon:2011gg} results in the sharpening of the pulse so that, eventually, 
it achieves sufficient concentration to form a black hole. The black hole critical behavior
discovered by Choptuik~~\cite{Choptuik:1992jv} in AF spacetimes appears here repeatedly. In particular,
after each reflection off the AdS boundary, there is yet another threshold for prompt BH formation (before
the next bounce).

To explain this behavior, a study of the normal modes of
scalar~\cite{Bizon:2011gg} and tensor~\cite{Dias:2011ss} perturbations revealed a non-linear mode
coupling at third order, shifting energy to higher frequencies. Consequently, this shift in frequency
would eventually lead to the formation of a black hole and so generic instability of AdS was conjectured. 
Interestingly though, further studies suggested the existence of non-linearly 
stable solutions~\cite{Dias:2012tq,Buchel:2012uh,Maliborski:2013jca,lieblingtalk}. These observations hint that a
straightforward application of the resonance picture can only partially capture the dynamical behavior.

We consider the dynamics of boson stars in AdS, motivated towards
better understanding of this instability and its implications for holographic scenarios. 
Horizon formation in gravitational collapse from the holographic
perspective implies thermalization of the dual conformal 
theory. When the gravitational problem is formulated in 
global $d+1$ space-time dimensional asymptotically
$AdS_{d+1}$, the dynamics of the dual $d$ space-time dimensional 
$CFT_{d}$ occurs on a $S^{d-1}$ sphere. Furthermore, the holographic duality 
arises as a correspondence between a String Theory and a CFT --- it 
can be truncated to a gravity/CFT duality when a CFT has a large number 
of strongly interacting degrees of freedom\footnote{This
can be quantified as a large central charge for even $d$,
or more generically, the large number of  excited degrees of freedom 
at thermal equilibrium.}. A natural expectation from the 
field theory perspective is that  a large number of strongly 
interacting degrees of freedom in a finite volume  thermalize
from generic initial conditions;
thus, one does expect no-threshold BH formation as advocated 
in~\cite{Bizon:2011gg}. Early work on boson stars in 
AdS \cite{Astefanesei:2003qy} demonstrated that such stationary 
configurations are linearly-stable. Thus, initializing a CFT state 
as a dual to an AdS boson star might result in a slow thermalization 
of the latter.

In this paper we show that, rather, boson stars 
in AdS are {\it non-linearly} stable. This non-linear stability
is unaffected by boson star perturbations, as long as these perturbations 
are sufficiently small. 
Even more surprising, we find that families of 
initial conditions with widely distributed mass-energy 
are non-linearly 
stable for sufficiently small mass as well. This suggests that 
non-linear stability is not a feature of states 
carrying a global charge (as dual to boson stars). As a result, 
it appears that there exist large sets of initial configurations 
in CFT which never thermalize in their evolution.      

The paper is organized as follows. In section~\ref{section2} 
we setup the gravitational dual of the generic, spatially-isotropic 
$CFT_3$, initial condition
specified by a pair of dimension-3 operators with a global 
$U(1)$ symmetry: an Einstein gravity with a negative cosmological 
constant and a massless, minimally coupled bulk complex scalar field.
In section~\ref{section3} we construct stationary configurations 
of the coupled scalar-gravity system, charged under 
the global $U(1)$ symmetry --- { boson stars}.
We show that boson stars are stable under 
linearized fluctuations. In section~\ref{section4} we present 
results of fully-nonlinear simulations of genuine boson stars, 
perturbed boson stars, and $U(1)$-neutral initial configurations 
with widely distributed, bulk mass-energy. 
We conclude in section~\ref{section5} and outline future directions.

\section{Effective action and equations of motion }\label{section2}
In this section, following \cite{Buchel:2012uh}, 
we review the formulation of the problem of the 
gravitational collapse of a complex scalar in 
asymptotically anti-de-Sitter space-time. 
We focus on asymptotically $AdS_4$ collapse, dual to 
$CFT_3$ (we choose $d=3$ in the notation of~\cite{Buchel:2012uh}).

The effective four-dimensional action is given by\footnote{We set the radius of AdS to 
one.}
\begin{equation}
\begin{split}
S_4=\frac{1}{16\pi G_4} \int_{\calm_4}d^4\xi \sqrt{-g}\left(R_4+6
-2\del_\mu\phi\del^\mu\phi^*\right)\,,
\end{split}
\eqlabel{ac4}
\end{equation}
where $\phi\equiv \phi_1+i\ \phi_2$ is a complex scalar field
and 
\begin{equation}
\calm_{4}=\del\calm_{3}\times \cali,\qquad \del\calm_{3}=R_t\times S^{2}\,,\qquad
\cali=\{x\in[0,\frac\pi 2]\}\,.
\eqlabel{manifold}
\end{equation}
Adopting the line element as in~\cite{Jalmuzna:2011qw},
\begin{equation}
ds^2 = \frac{1}{\cos^2 x} \left(
                                     -Ae^{-2\delta} dt^2
                                     +\frac{dx^2}{A}
                                     +\sin^2x \, d\Omega^2_{2}
                                        \right) \, ,
\eqlabel{eq:metric}
\end{equation}
where 
$d\Omega^2_{2}$ is the metric of unit radius 
$S^{2}$, and $A(x,t)$ and $\delta(x,t)$ 
are scalar functions describing the metric. Rescaling  the
matter fields as in \cite{Buchel:2012uh}
\begin{eqnarray}
\hat \phi_i & \equiv  & \frac{\phi_i}{\cos^{2}x} \, , \\
\hat \Pi_i  & \equiv  & \frac{e^\delta}{A}\frac{\partial_t \phi_i}{\cos^{2}x} 
                 \, ,\label{pihdef}\\ 
\hat \Phi_i & \equiv  & \frac{\partial_x\phi_i}{\cos x}
                  \, ,
\end{eqnarray}
we find the following equations of motion 
(we drop the caret from here forward)
\begin{equation}
\begin{split}
\dot \phi_i  = & A e^{-\delta} \Pi_i\,, \\
\dot \Phi_i  = & \frac{1}{\cos x} \left( \cos^{2}x A e^{-\delta} \Pi_i \right)_{,x}\,,\\
\dot \Pi_i   = & \frac{1}{\sin^{2}x} \left( \frac{\sin^{2}x}{\cos x}
                                               A e^{-\delta} \Phi_i
                                               \right)_{,x}\,.
\end{split}
\eqlabel{kg}
\end{equation}
\begin{equation}
\begin{split}
A_{,x}       = &   \frac{1 + 2 \sin^2x}{\sin x \cos x} \left(1-A\right)
                  - \sin x \cos^{5}x A\left( \frac{\Phi_i^2}{\cos^2x} + \Pi_i^2 \right)\,,\\
\delta_{,x}  = &  - \sin x \cos^{5}x \left( \frac{\Phi_i^2}{\cos^2x}+ \Pi_i^2 \right)\,,
\end{split}
\eqlabel{constx}
\end{equation}
together with one constraint equation
\begin{equation}
A_{,t} + 2 \sin x \cos^{4}x A^2 e^{-\delta} \left( \Phi_i \Pi_i \right)=0\,,
\eqlabel{consteq}
\end{equation}
where a sum over $i= \{1,2\}$ is implied.
We are interested in studying the solution to \eqref{kg}-\eqref{consteq}
subject to the boundary conditions:  
\nxt Regularity at the origin implies these quantities behave  as
\begin{equation}
\begin{split}
\phi_i(t,x)  = & \phi_0^{(i)}(t) + {\cal O}(x^2)\,,  \\
A(t,x)       = &  1 + {\cal O}(x^2)\,,  \\
\delta(t,x)  = &  \delta_0(t) + {\cal O}(x^2)\,;
\end{split}
\eqlabel{ir}
\end{equation}
\nxt at the outer boundary $x=\pi/2$ we introduce 
$\rho \equiv \pi/2-x$ so that we have
\begin{equation}
\begin{split}
\phi_i(t,\rho)  = & \phi_3^{(i)}(t)\rho + {\cal O}(\rho^3)\,,  \\
A(t,\rho)       = &  1 - M \frac{\sin^3\rho}{\cos \r} 
+  {\cal O}(\rho^{6})\,,  \\
\delta(t,\rho)  = &  0 + {\cal O}(\rho^{6})\,.
\end{split}
\eqlabel{uvt}
\end{equation}

The asymptotic behavior \eqref{uvt} determines the boundary CFT observables: 
the expectation values of the stress-energy tensor $T_{kl}$, and the
operators $\calo_{3}^{(i)}$, dual to $\phi_i$, 
\begin{equation}
\begin{split}
&8\pi G_{4}\langle T_{tt}\rangle =M\,,\qquad 
\langle T_{\a\b}\rangle= \frac{g_{\a\b}}{2}\ \langle T_{tt}\rangle\,,\\
&16\pi G_{d+1}\langle \calo_{3}^{(i)} \rangle=12\ \phi_3^{(i)}(t)\,,
\end{split}
\eqlabel{vevs}
\end{equation}
where  $g_{\a\b}$ is a metric on a round $S^{2}$. 
Additionally note that the conserved $U(1)$ charge is given by 
\begin{equation}
Q=8\pi\ \int_0^{\pi/2} dx\ 
\sin^{2}x\ \cos^{2}x\ \left(\Pi_2(0,x) \phi_1(0,x)-\Pi_1(0,x)\phi_2(0,x)\right)  ,
\eqlabel{qinit}
\end{equation} 
and that since $\del_t Q=0$, the integral in \eqref{qinit} can be evaluated at 
$t=0$.

The constraint \eqref{consteq} implies that $M$ in \eqref{uvt} 
is  time-independent, ensuring            energy
conservation
\begin{equation}
\del_t\ \langle T_{tt}\rangle=0\,.
\eqlabel{emc} 
\end{equation}

It is convenient to introduce the mass aspect function $\calm(t,x)$
as 
\begin{equation}
A(t,x)=1 - \calm(t,x) \frac{\cos^3 x}{\sin x}\, .
\eqlabel{massfunction}
\end{equation}
Following \eqref{constx} we find 
\begin{equation}
\calm(t,x)=\int_0^x dz\ \tan^{2} z\ \cos^{4}z\  A(t,z) 
\left[\frac{\Phi_i^2(t,z)}{\cos^2 z}+\Pi_i^2(t,z)\right]\,.
\eqlabel{mf1}
\end{equation}
Comparing \eqref{mf1} and \eqref{uvt} we see that
\begin{equation}
M=\calm(t,x)\bigg|_{x=\frac\pi2}\,.
\eqlabel{mcalm}
\end{equation}

 \section{Boson stars in $AdS_4$}\label{section3}
There is an interesting class of stationary,
perturbatively stable, fully-nonlinear solutions to \eqref{kg}-\eqref{uvt}
with nonzero $Q$, referred to as {\it boson stars}
~\cite{Astefanesei:2003qy,2012LRR....15....6L}. 
Such solutions are characterized by a discrete integer $n=0,1,\cdots$,
denoting the number of nodes of the complex scalar radial 
wave-function, and a continuous value of the global charge $Q$. 
In this section we discuss the numerical construction of such solutions,
their perturbative properties, linearized stability, 
and their relation (for small $Q$) 
to linearized $AdS_4$ massless, minimally coupled, scalar 
modes --- the {\it oscillons}. Oscillons were reviewed 
in detail in~\cite{Buchel:2012uh}. These stationary solutions 
are uniquely characterized 
by an excitation level $j=\{0,1,\cdots\}$:
\begin{equation}
\begin{split}
&e_{j}(x)= d_j\ \cos^3 x\ _2F_1\left(-j,3+j;\frac 32;\sin^2 x\right)\,,\\
&w^{(j)}=3+2j\,,\qquad d_j=\left(\frac{16 (j+1)(j+2)}{\pi}\right)^{1/2}\,,
\end{split}
\eqlabel{defoscil}
\end{equation}   
where $w^{(j)}$ is an oscillon frequency and $d_j$ is a  
constant enforcing their orthonormality,
\begin{equation}
\int_0^{\pi/2} dx\ e_i(x)e_j(x) \tan^{2} x=\dd_{ij}\,.
\eqlabel{os2a}
\end{equation}

\subsection{Stationary boson stars}

Assuming a stationary solution in which the complex field varies harmonically
\begin{equation}
\phi_1(x,t)+i \phi_2(x,t)=\frac{\phi(x)}{\cos^2 x} e^{i\w t}\,,\qquad A(t,x)=a(x)\,,\qquad \dd(t,x)=d(x)\,, 
\eqlabel{statdef}
\end{equation}
we find a system of ODEs from \eqref{kg} and \eqref{constx}
\begin{equation}
\begin{split}
0=&\phi''+\left(\frac{2}{\cos x \sin x}+\frac{a'}{a}-d'\right)\  \phi'+ \omega^2 e^{2 d}a^{-2}\ \phi\,,\\
0=&d'+ \sin x \cos x\ a^{-2}\left((\phi')^2 a^2+\phi^2 \omega^2 e^{2 d}\right)\,,\\
0=&a'+\frac{2 \cos^2 x-3}{\cos x \sin x} (1-a)+ \sin x \cos x\ a^{-1}
\left((\phi')^2 a^2+\phi^2 \omega^2 e^{2 d}\right)\,.
\end{split}
\eqlabel{steoms}
\end{equation}
The charge and the mass determined by these solutions are given by 
\begin{equation}
\begin{split}
Q=&8\pi\ \int_0^{\pi/2}\ dx\ \frac{\w \sin^2 x \phi(x)^2 e^{d(x)}}{a(x) \cos^2 x }\,,\\
M=&\int_0^{\pi/2}\ dx\ \frac{ \sin^2 x }{a(x) \cos^2 x }\left(a(x)^2 (\phi'(x))^2+e^{2d(x)}\w^2\phi(x)^2\right)\,.
\end{split}
\eqlabel{chargemassb}
\end{equation}

A physically relevant solution to equations~\ref{steoms} must satisfy:
\nxt At the origin of AdS, \ie,  $x\to 0$: 
\begin{equation}
\begin{split}
a=&1-\frac13 (p_0^h)^2 \w^2 e^{2 d_0^h}\ x^2+\calo(x^4)\,,\\
d=&d_0^h-\frac12 (p_0^h)^2 \w^2 e^{2 d_0^h}\ x^2+\calo(x^4)\,,\\
\phi=&p_0^h-\frac16 p_0^h \w^2 e^{2 d_0^h}\ x^2+\calo(x^4)\,.
\end{split}
\eqlabel{irpert}
\end{equation}
Note that besides $\w$,  the general solution is characterized by
\begin{equation}
\{p_0^h\,,\ d_0^h\}\,.
\eqlabel{irpars}
\end{equation}
\nxt and asymptotically (at the AdS boundary, \ie, $\r\to 0$): 
\begin{equation}
\begin{split}
a=&1+a_3^b\ \r^3+\calo(\r^6)\,,\\
d=&\frac32 (p_3^b)^2 \r^6+\left(\frac34 (p_3^b)^2-\frac14 (p_3^b)^2 \omega^2\right) \r^8+\calo(\r^9)\,,\\
\phi=&p_3^b \r^3+\left(\frac25 p_3^b-\frac{1}{10} p_3^b \omega^2\right) \r^5+\calo(\r^6)\,.
\end{split}
\eqlabel{uv}
\end{equation}
Note that besides $\w$,  the general solution is characterized by
\begin{equation}
\{p_3^b\,,\ a_3^b\}\,.
\eqlabel{uvpars}
\end{equation}

From \eqref{irpars} and \eqref{uvpars} we have precisely the correct number of coefficients to 
find an isolated solution\footnote{As we discuss in section \ref{pertbs} solutions of 
a fixed charge are labeled by an integer, specifying the ``level'' of a boson 
star.} for a given $\w$.

\subsection{Small charge boson stars as oscillons}\label{pertbs}
In this section we discuss analytic results for the spectrum of boson stars perturbatively 
in the amplitude $\phi$. To this end we introduce 
\begin{equation}
\begin{split}
&\phi=\l\ \phi_1+\calo(\l^3)\,,\qquad a=1+\l^2\ a_2+\calo(\l^4)\,,\\
&e^d=1+\l^2\ d_2+\calo(\l^4)\,,\qquad \w=\w_0+\w_2\l^2+\calo(\l^4)\,,
\end{split}
\eqlabel{pert}
\end{equation}
where $\l$ is an expansion parameter. 
Substituting \eqref{pert} into \eqref{steoms} we find that the equation for $\phi_1$ decouples 
to leading order
\begin{equation}
\begin{split}
0=&\phi_1''+\frac{2}{\sin x\cos x}\ \phi_1'+\w_0^2\ \phi_1\,.
\end{split}
\eqlabel{perlead}
\end{equation}
Normalizing $\phi_1$ as 
\begin{equation}
\phi_1\bigg|_{x\to 0_+}=1 \, ,
\end{equation}
the general solution of \eqref{perlead}, subject to the boundary 
conditions \eqref{irpert} and \eqref{uv}, is given by 
\begin{equation}
\begin{split}
\phi_1^{(j)}=\frac{1}{d_j}\ e_j(x)\,,
\qquad \w_0^{(j)}=w^{(j)}=3+2j \,,
\end{split}
\eqlabel{solvelead}
\end{equation}
where the integer $j=0,1,2\cdots$ parameterize the 
'excitation level' of the boson star. Notice that
the asymptotic expansion for $\phi$ (in eqn~\ref{irpert}) implies 
$p_0^h=\lambda$ .
Given \eqref{solvelead},
we find from \eqref{chargemassb}
\begin{equation}
\begin{split}
&M=\l^2\ \frac{ \pi (3+2j)^2}{8(j+1)(j+2)}+\calo(\l^4)\,,
\qquad Q=\l^2\ \frac{ \pi^2(3+2j)}{2(j+1)(j+2)}+\calo(\l^4)\,,\\
&M=\frac{3+2j}{4\pi}\ Q+ \calo(Q^2)=\frac{\w_0^{(j)}}{4\pi}\ Q+\calo(Q^2)\,.
\end{split}
\eqlabel{mql}
\end{equation}  
Note from \eqref{mql} that for a fixed charge $Q$, excited levels 
of boson stars are more massive.

It is straightforward to compute the leading-order background 
warp factors $\{a_2,d_2\}$, as well as subleading frequency 
correction $\w_2$. 
In what follows we present explicit expressions for the first 
four levels of a boson star:
\nxt $j=0$ level, 
\begin{equation}
\begin{split}
a_2^{(0)}=\frac{9\cos^3 x}{8\sin x}\ \left(\frac 14\sin(4x)-x \right)\,,
\end{split}
\eqlabel{a20}
\end{equation}
\begin{equation}
\begin{split}
d_2^{(0)}=\frac 32 \cos^6 x\,,
\end{split}
\eqlabel{d20}
\end{equation}
\begin{equation}
\begin{split}
\w_2^{(0)}=-\frac{63}{32}\,;
\end{split}
\eqlabel{w20}
\end{equation}
\nxt $j=1$ level, 
\begin{equation}
\begin{split}
a_2^{(1)}=\frac{25\cos^3 x}{72\sin x}\left(\frac 14 
\sin(4x)(2\cos(4x)+1)-3x\right)\,,
\end{split}
\eqlabel{a21}
\end{equation}
\begin{equation}
\begin{split}
d_2^{(1)}=\frac{5}{18}\cos^6 x\left(32 \cos^4 x -40 \cos^2 x+15\right)\,,
\end{split}
\eqlabel{d21}
\end{equation}
\begin{equation}
\begin{split}
\w_2^{(1)}=-\frac{3025}{864}\,;
\end{split}
\eqlabel{w21}
\end{equation}
\nxt $j=2$ level,
\begin{equation}
\begin{split}
a_{2}^{(2)}=\frac{49\cos^3 x}{144\sin x}\left(\frac 18\sin(8x) (2 \cos(4x)+1)-3x\right)\,,
\end{split}
\eqlabel{a22}
\end{equation}
\begin{equation}
\begin{split}
d_2^{(2)}=\frac{7}{90}\cos^6 x\left(960 \cos^8 x -2240 \cos^6 x+1904\cos^4 x-700\cos^2x+105\right)\,,
\end{split}
\eqlabel{d22}
\end{equation}
\begin{equation}
\begin{split}
w_2^{(2)}=-\frac{89327}{17280}\,;
\end{split}
\eqlabel{w22}
\end{equation}
\nxt $j=3$ level, 
\begin{equation}
\begin{split}
a_2^{(3)}=\frac{81\cos^3 x }{400\sin x}\left(\frac 18\sin(8x)(2\cos(8x)+2 \cos(4x)+1)-5x\right)\,,
\end{split}
\eqlabel{a23}
\end{equation}
\begin{equation}
\begin{split}
d_2^{(3)}=&\frac{9}{350} \cos^6x (28672 \cos^{12}x-96768 \cos^{10}x+130752 \cos^8x
-90048 \cos^6x\\
&+33264 \cos^4x-6300 \cos^2x+525)\,,
\end{split}
\eqlabel{d23}
\end{equation}
\begin{equation}
\begin{split}
\w_2^{(3)}=-\frac{154143}{22400}\,.
\end{split}
\eqlabel{w23}
\end{equation}
Further comparing the above with \eqref{irpert} and \eqref{uv}, we identify
the condensates perturbatively in  $p_0^h=\l$ (see Table~\ref{table1}).

\begin{table}
\begin{center}
\begin{tabular}{|c|c|c|c|c|}
	\hline
$j$ &  $\frac{\w-\w_0^{(j)}}{\l^2}+\calo(\l^2)$    &   $\frac{d_0^h}{\l^2}+\calo(\l^2)$  & $\frac{p_3^b}{\l}+\calo(\l^2)$   
& $\frac{a_3^b}{\l^2}+\calo(\l^2)$  \\
	\hline
0   & $-\frac{63}{32}$ & $\frac32$ & $1$  & $-\frac{9}{16}\pi$ \\
& & & &\\
1   & $-\frac{3025}{864}$ & $\frac{35}{18}$ & $-\frac 53$  
& $-\frac{25}{48}\pi$\\
& & & & \\
2   & $-\frac{89327}{17280}$ & $\frac{203}{90}$ 
& $\frac73$ & $-\frac{49}{96}\pi$ \\
& & & &\\
3   & $-\frac{154143}{22400}$ & $\frac{873}{350}$ & $-3$  
& $-\frac{81}{160}\pi$\\
	\hline
\end{tabular} 
\end{center}
\caption{Condensate values \eqref{irpert} 
and \eqref{uv} of boson stars,
perturbatively in  $p_0^h=\l$. }
\label{table1}
\end{table}

\subsection{Numerical boson stars}
\begin{figure}[t]
\begin{center}
\psfrag{p3}{$p_3^{b}$}
\psfrag{a3}{$a_3^{b}$}
\psfrag{p0}{$p_0^h$}
\psfrag{dd0}{$d_0^h$}
\psfrag{om}{$(\w^{(j)}/\w_0^{(j)})^2$}
  \includegraphics[width=2.5in]{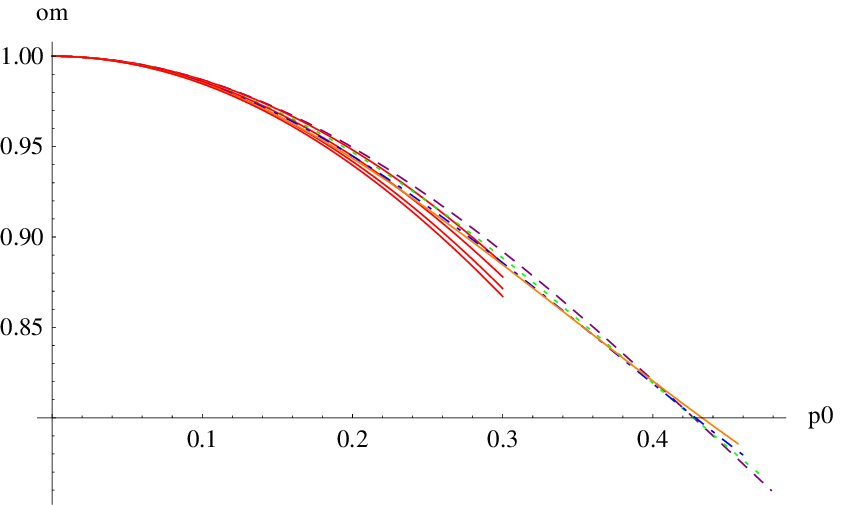}
  \includegraphics[width=2.5in]{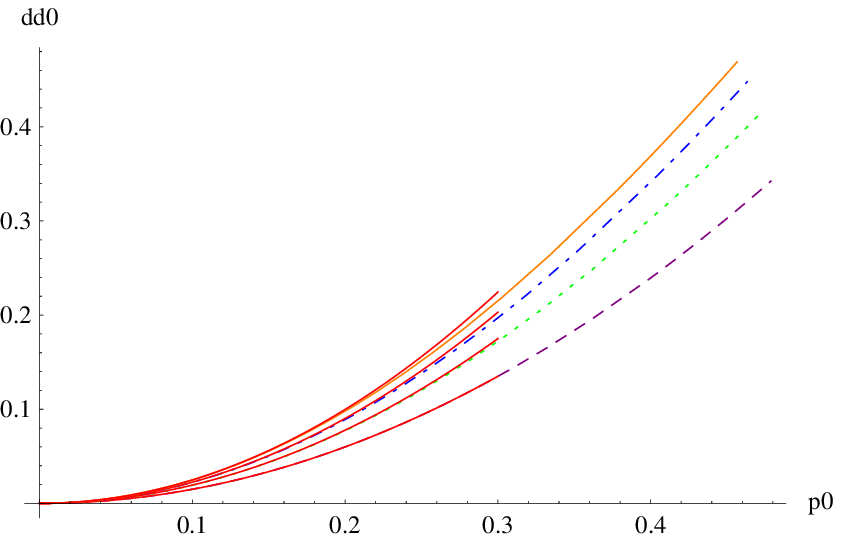}
  \includegraphics[width=2.5in]{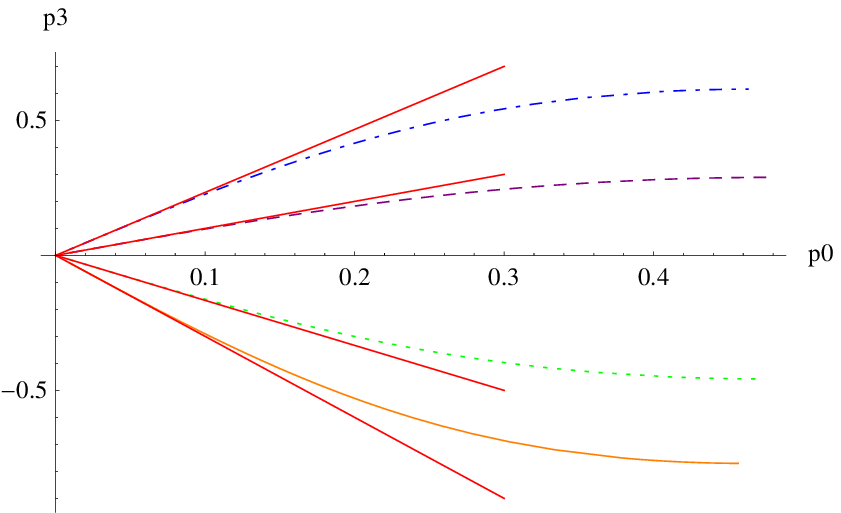}
  \includegraphics[width=2.5in]{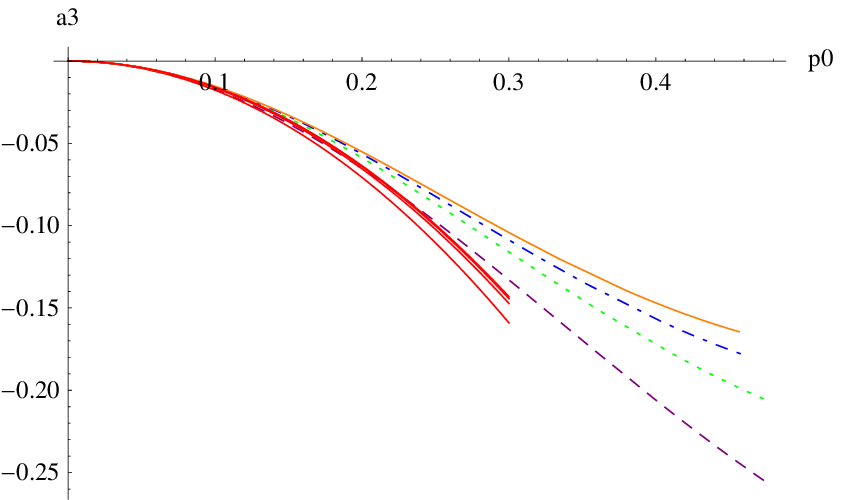}
\end{center}
  \caption{(Colour online) Condensate values \eqref{irpert} and \eqref{uv} for level $j=\{0,1,2,3\}$ 
 ( $\{$dashed purple, dotted green,dot-dashed blue, solid orange$\}$ curves) boson stars. The red lines represent perturbative in $p_0^h$
approximations, see Table \ref{table1}.   
}
\label{figure1}
\end{figure}

\begin{figure}[t]
\begin{center}
\psfrag{MM}{$M$}
\psfrag{Q}{$Q$}
\psfrag{p0h}{$p_0^h$}
  \includegraphics[width=2.5in]{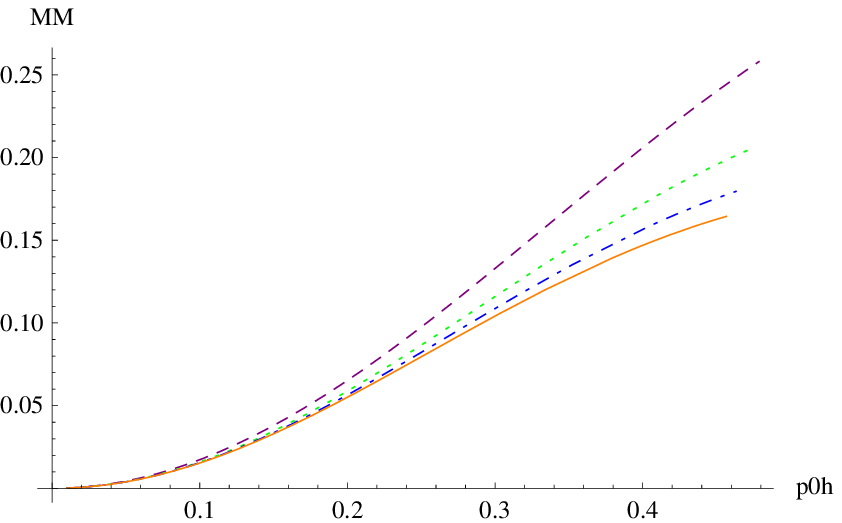}
  \includegraphics[width=2.5in]{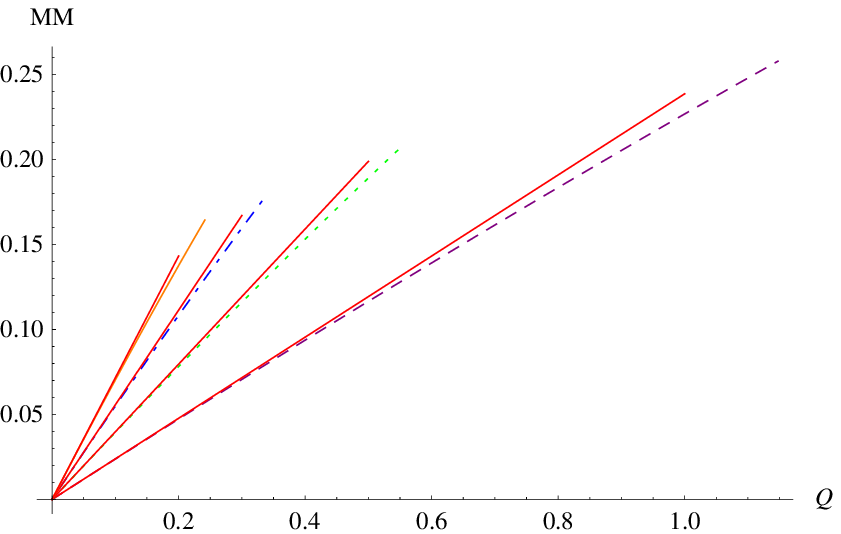}
\end{center}
  \caption{(Colour online) Mass vs. $p_0^h$ (left panel) and vs. charge (right panel) 
for level $j=\{0,1,2,3\}$ 
 ( $\{$dashed purple, dotted green,dot-dashed blue, solid orange$\}$ curves) boson stars.  
The solid red lines represent perturbative in $Q$
approximations, see \eqref{mql}.   
}
\label{figure2}
\end{figure}

In the previous section we identified the first four levels of a boson star
perturbatively in the amplitude $p_0^h$. Here we report results 
for $\{\w,d_0^h,p_3^b,d_3^b\}$, as well as $\{M,Q\}$, for generic $p_0^h$
from the numerical solution  of \eqref{steoms}.
These results are collected in Figs.~\ref{figure1} and \ref{figure2}. 

The top-left panel of Fig.~\ref{figure1} presents 
the frequency $\w^{(j)}$ of a level-$j$ 
boson star rescaled  to that of a level-$j$  oscillon 
frequency $\w_0^{(j)}$ (see \eqref{solvelead}). 
We use purple/green/blue/orange color coding to denote 
$j=0\cdots 3$. As $p_0^h$ 
(and correspondingly the mass and the charge --- see Fig.~\ref{figure2} ) 
of a boson star grows, its frequency decreases. The remaining panels in
Fig.~\ref{figure1} present the dependence of $\{d_0^h,p_3^b,d_3^b\}$ as a 
function of $p_0^h$. Notice that $p_3^b$ saturates; this saturation is the main 
obstacle in generating boson stars with ever increasing values of $p_0^h$ (or mass).    
The red curves indicate perturbative approximations in $p_0^h$ as collected 
in Table~\ref{table1}, (Fig.~\ref{figure1}), and perturbative approximation \eqref{mql} in 
 $Q$ (right panel of Fig.~\ref{figure2}).

\begin{figure}[t]
\begin{center}
\psfrag{m}{$\ln (i+1)$}
\psfrag{p}{$\ln c_i^{(j)}$}
  \includegraphics[width=4in]{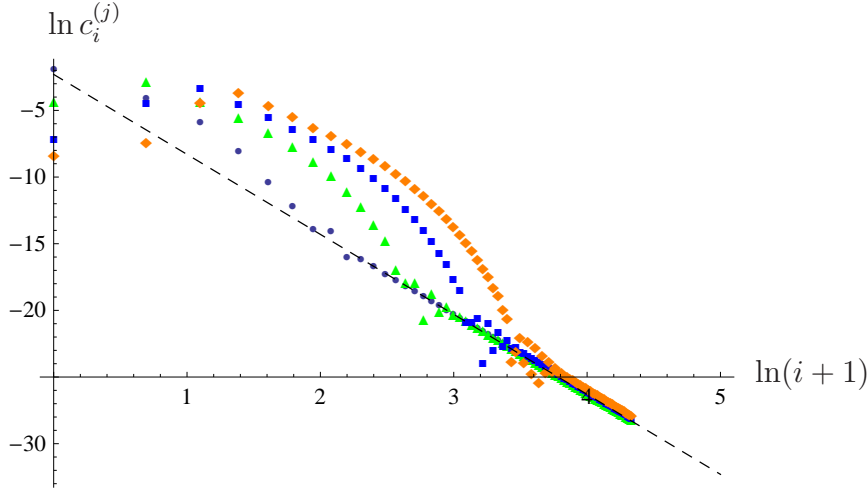}
\end{center}
  \caption{(Colour online) Spectral decomposition of 
level $j=\{0,1,2,3\}$ ($\{$purple circles,green triangles,blue squares,orange diamonds$\}$)
boson stars in oscillon basis, see \eqref{os3}. For large 
$i$ all the curves approach a universal fall-off $c_i^{(j)}\propto (1+i)^{-6}$
(the black dashed curve).   
}
\label{figure3a}
\end{figure}
In the limit of vanishing charge $Q$, 
the level-$j$ boson star radial profile $\phi^{(j)}$ is 
a single level-$j$ oscillon
(see \eqref{solvelead}, \eqref{mql}) :
\begin{equation}
\phi^{(j)}(x)\propto \sqrt{Q}\ e_j(x)\,.
\eqlabel{os1}
\end{equation}
For finite $Q$ all the oscillons are excited. In Fig.~\ref{figure3a} 
we present the spectral decomposition in the oscillon basis of the most massive level-$j$ boson stars that we were able to construct
\begin{equation}
c_i^{(j)}\equiv \bigg|\ \int_0^{\pi/2} dx\ \phi^{(j)}(x) e_i(x) \tan^2 x\ \bigg|\,.
\eqlabel{os2}
\end{equation}
Note that the maxima of $c_i^{(j)}$ are achieved for $i=j$, much like in the small-$Q$ limit.
For all levels considered $c_i^{(j)}$ approach a universal fall-off:
\begin{equation}
c_i^{(j)}\ \propto\ (1+i)^{-6}\,,\qquad i\gg j\,,
\eqlabel{os3}
\end{equation}
represented by a dashed black curve in Fig.~\ref{figure3a}.

\subsection{Perturbative stability of boson stars}
In this section we explore the linearized stability of boson stars. 
Consider perturbations of stationary solutions \eqref{statdef} to leading 
order in $\e$:
\begin{equation}
\begin{split}
&\phi_1(x,t)+i \phi_2(x,t)={\cos^{-2} x} \biggl(
\phi(x)+\e (f_1(t,x)-i \phi(x) g_1(t,x))\biggr)e^{i\w t}\,,\\
&A(t,x)=a(x)+\e\ a_1(t,x)\,,\\
&\dd(t,x)=d(x)+\e\ \dd_1(t,x) \,.
\end{split}
\eqlabel{fluceoms}
\end{equation} 
Further introducing 
\begin{equation}
\begin{split}
f_1(t,x)=F_1(x)\cos(\c t)\,,\qquad g_1(t,x)=-G_1(x)\ \sin(\c t)\,,
\end{split}
\eqlabel{furtheransatz}
\end{equation}
the equations for $a_1(t,x)$ and $\dd_1(t,x)$ can be  solved explicitly:
\begin{equation}
a_1(t,x)=\sin(2x) a(x)\left(\w\ \phi(x)^2\ G_1'(x)-\phi'(x)\ F_1(x)\right)\ \cos(\c t)\,,
\eqlabel{a1solve}
\end{equation}
\begin{equation}
\begin{split}
&\dd_1(t,x)=-\frac{e^{-2 d(x)}}{\cos(x) \phi(x) \omega \sin(x)} 
\biggl(a(x)^2 \cos(x) \phi(x) G_1''(x) \sin(x)+(\\
&-2 \cos(x)^2 \phi(x)^3 e^{2 d(x)} \sin(x)^2 \w^2+a(x) (2 a(x) 
\cos(x) \phi'(x) \sin(x)\\
&+2 a(x) \cos(x)^2 \phi(x)-a(x) \phi(x)-2 \cos(x)^2 \phi(x)+3 \phi(x)))\ G_1'(x)
\\
&+\cos(x) \phi(x) e^{2 d(x)}\ G_1(x) \chi^2 \sin(x)-2 \cos(x) F(x) (\cos(x)^3 \phi(x) 
\phi'(x)\\
&-\cos(x) \phi(x) \phi'(x)-\sin(x)) \w e^{2 d(x)}\biggr)\  \cos(\chi t)\,.
\end{split}
\eqlabel{d1solve}
\end{equation}
where $F_1(x)$ and $G_1(x)$ satisfy a coupled system of equations
\begin{equation}
\begin{split}
&0=F_1''+\frac{-2 \cos(x)^2+2 \cos(x)^2 a+3-a}{a \sin(x) \cos(x)}\  F_1'
-2 \phi \omega\ G_1''+\frac{2 \w}{a \sin(x) \cos(x)} \biggl(\\
&2 \sin(x) \phi^2 \phi' \cos(x)^3+2 \cos(x)^2 \phi-2 a \phi \cos(x)^2-2 \sin(x) 
a \phi' \cos(x)\\
&-3 \sin(x) \phi^2 \phi' \cos(x)+a \phi-3 \phi\biggr)\ G_1'
-\biggl(4 a \cos(x)^2 (\phi')^2-e^{2d} \chi^2\\
&-6 (\phi')^2 a+3 e^{2d} \w^2\biggr) a^{-2}\ F_1\,,
\end{split}
\eqlabel{fgsystem1}
\end{equation}
\begin{equation}
\begin{split}
&0=G_1'''+\biggl(-6 \cos(x)^2 \phi+6 a \phi \cos(x)^2+2 \sin(x) a \phi' \cos(x)
+9 \phi\\
&-7 a \phi\biggr)\biggl(a \phi \sin(x) \cos(x)\biggr)^{-1}\ G_1''
+\biggl(5 a^3 \phi^2+4 a^3 \sin(x) \phi' \phi \cos(x)^3\\
&-10 a^3 \sin(x) \phi' \phi \cos(x)-4 a^2 \sin(x) \phi' \phi \cos(x)^3
+6 a^2 \sin(x) \phi' \phi \cos(x)\\
&+2 e^{2d} a \phi^2 \w^2 \cos(x)^4-e^{2d} a \cos(x)^4 \phi^2 \c^2
-2 e^{2d} a \phi^2 \w^2 \cos(x)^2\\
&+e^{2d} a \cos(x)^2 \phi^2 \c^2
+2 a^2 (\phi')^2 \cos(x)^6 \phi^2-5 a^2 (\phi')^2 \cos(x)^4 \phi^2\\
&+3 a^2 (\phi')^2 \cos(x)^2 \phi^2
+9 e^{2d} \cos(x)^2 \phi^4 \w^2
+6 e^{2d} \cos(x)^6 \phi^4 \w^2\\
&-15 e^{2d} \cos(x)^4 \phi^4 \w^2-16 a^3 \phi^2 \cos(x)^2
+8 a^3 \phi^2 \cos(x)^4+9 a \phi^2-12 a^2 \phi^2\\
&+24 a^2 \cos(x)^2 \phi^2-12 a^2 \cos(x)^4 \phi^2
-12 a \cos(x)^2 \phi^2+4 a \cos(x)^4 \phi^2\\
&+2 a^3 (\phi')^2 \cos(x)^4
-2 a^3 (\phi')^2 \cos(x)^2\biggr)\biggl(a^3 \phi^2 \sin(x)^2 \cos(x)^2\biggr)^{-1}\ G_1'
\\&+\frac{2 e^{2d} \w}{a^2 \phi}\ F_1'+(4 \phi^2 \cos(x)^2-2 a-6 \phi^2) e^{2d} 
 \w \phi'\biggl(a^3 \phi^2\biggr)^{-1}\ F_1\,.
\end{split}
\eqlabel{fgsystem2}
\end{equation}
Notice that \eqref{fgsystem1}-\eqref{fgsystem2} are left invariant 
under the shift
\begin{equation}
G_1\to G_1+\calg\,,
\eqlabel{g1shift}
\end{equation}
where $\calg$ is an arbitrary constant. From \eqref{d1solve} is it 
clear that this constant is fixed uniquely requiring that 
\begin{equation}
\lim_{x\to \pi/2}\ \dd_1(t,x)=0\,,
\eqlabel{d1zero}
\end{equation}
\ie, we keep the time coordinate at the boundary fixed.
We do not have to worry about the shift symmetry \eqref{g1shift},
provided we rewrite  \eqref{fgsystem1}-\eqref{fgsystem2} using 
\begin{equation}
dG_1(x)\equiv G_1'(x) \,.
\eqlabel{dG1def}
\end{equation}

Eqs.~\eqref{fgsystem1}-\eqref{fgsystem2} must be solved subject to 
constraints that $F_1(x)$ and $G_1(x) \phi(x)$ are 
regular for $x\in [0,\pi/2)$, and have only normalizable 
modes as $x\to \pi/2$, \ie,
\begin{equation}
F_1\propto \r^3\,,\qquad G_1\propto {\rm const}\,,\qquad {\rm as}
\qquad \r\to 0\,.
\eqlabel{normalizable}
\end{equation}
The latter regularity condition implies that 
$G_1$ can have a simple pole (or $dG_1$ can have a double pole)
precisely where $\phi(x)$ has a zero\footnote{Recall that excited 
level stationary boson stars are characterized by the number of nodes 
in the radial profile $\phi(x)$, see \eqref{statdef}.}. 
These poles  represent a technical difficulty in identifying 
the fluctuations about excited boson stars 
---  specifically, a straightforward shooting method: 
integrating from both boundaries with suitable boundary conditions and demanding
continuity at an arbitrary radial location
will invariably encounter these poles rendering this method delicate to apply.

Here we discuss the fluctuations about the 
lowest-level (ground state) boson stars, and also present 
analytic results for fluctuations about perturbatively-light 
excited boson stars\footnote{We verified explicitly that while $dG_1^{(1)}$  
to order $\calo(\l^3)$ has a double pole at the location of the zero 
of $\phi^{(1)}$  (constructed perturbatively in $\l$
to order $\calo(\l^5)$ inclusive --- see \eqref{pert}),
the full radial profile of physical fluctuations,  $G_1^{(1)} \phi^{(1)}$, is 
smooth for $x\in[0,\pi/2]$.}.

Finally, since the system of equations \eqref{fgsystem1}-\eqref{fgsystem2}
is linear, we can further fix the normalizable mode $F_1$: 
\begin{equation}
\lim_{\r\to 0}\ \frac{F_1}{\r^3}=1\,.
\eqlabel{f1fix}
\end{equation}

Given \eqref{normalizable} and \eqref{f1fix}, we can specify 
the boundary conditions for $\{dG_1,F_1\}$:
\nxt at the origin of AdS, \ie, as $x\to 0$,
\begin{equation}
\begin{split}
dG_1=&2 g_1^h x +\calo(x^3)\,,\qquad F_1=f_0^h+\calo(x^2)\,;
\end{split}
\eqlabel{forigin}
\end{equation}
\nxt at the AdS boundary, \ie, as $\r\to 0$
\begin{equation}
\begin{split}
dG_1=&2 g_2^b \r +\calo(\r^3)\,,\qquad F_1=\r^3+\calo(\r^5)\,.
\end{split}
\eqlabel{fboundary}
\end{equation}

Note that along with $\chi$, the physical solution $\{dG_1,F_1\}$ 
is characterized 
by $\{g_1^h,f_0^h,g_2^b\}$ --- which is the correct number of coefficients 
necessary to 
determine a unique (or isolated) solution for a pair of coupled 
second-order ODEs   \eqref{fgsystem1}-\eqref{fgsystem2}.

Using the boundary conditions \eqref{forigin} and \eqref{fboundary}
it is easy to see that 
 the charge of a fluctuating boson star 
does not change to leading order in $\epsilon$:
\begin{equation}
\dd Q\ \propto\ \int_0^{\pi/2} dx\ \frac{d}{dx}\biggl\{
\frac{\sin(x)^2 \phi(x)^2 a(x) G_1'(x)}{e^{d(x)}\cos(x)^2}\biggr\}=0\,.
\eqlabel{deltaQ}
\end{equation}

\subsubsection{Linearized fluctuations about light boson stars}

We report here the results for solving 
\eqref{fgsystem1}-\eqref{fgsystem2} for light boson stars, 
\ie, perturbatively in $\l$, see section \ref{pertbs}. 
In general, we search solution to above equations 
as a series 
\begin{equation}
\begin{split}
&dG_1^{(j)}(x)=\l\ dG_{1,1}^{(j)}+\l^3\ dG_{1,3}^{(j)}+\calo(\l^5) \, , \\
&F_1^{(j)}=F_{1.0}^{(j)}
+\l^2\ F_{1,2}^{(j)}+\l^4\ F_{1,4}^{(j)}+\calo(\l^6) \, , \\
&\c^{(j)}=\c_0^{(j)}+\l^2\ \c_1^{(j)}+\l^4\ \c_2^{(j)}+\calo(\l^6) \, ;
\end{split}
\eqlabel{seriesg1f1}
\end{equation}
where $j$ is the excitation level of a boson star.

We find\footnote{For excited levels we present only the  
coefficients $\{\c_i^{(j)}\}\,,\ i=0,1\,,\ j=1\cdots 3$.}:
\nxt $j=0$ level,
\begin{equation}
\c_{0}^{(0)}=6\,,\qquad \c_1^{(0)}=-\frac{135}{32},\qquad 
\c_2^{(0)}=\frac{1215}{128}\pi^2-\frac{113892831}{1254400}\,,
\eqlabel{cj0}
\end{equation}
\begin{equation}
F_{1,0}^{(0)}=\cos^3 x\,,
\eqlabel{fj0}
\end{equation}
\begin{equation}
\begin{split}
&F_{1,2}^{(0)}=\frac{\cos(x)^3}{4480\sin(x)}  (5760 \cos(x)^8 \sin(x)
-14896 \cos(x)^6 \sin(x)\\
&+18738 \cos(x)^4 \sin(x)+7560 \cos(x)^3 x
-9153 \cos(x)^2 \sin(x)\\
&-15120 x \cos(x)+1890 \sin(x) \pi^2-7560 x^2 \sin(x))
-2 \cos(x)^5 (32 \cos(x)^4\\
&-54 \cos(x)^2+27)\calc_{1,1}\,,
\end{split}
\eqlabel{f2j0}
\end{equation}
\begin{equation}
\begin{split}
dG_{1,1}^{(0)}=& 
-\frac{1}{4480} \sin(x) \cos(x) (5760 \cos(x)^6-9376 \cos(x)^4+9900 
\cos(x)^2\\
&-2799)
+2 \sin(x) \cos(x) (8 \cos(x)^2-3) (4 \cos(x)^2-3) \calc_{1,1}\,.
\end{split}
\eqlabel{gj0}
\end{equation}
The integration constant $ \calc_{1,1}$ is not fixed at order $\calo(\l^2)$,
but is uniquely determined\footnote{This pattern 
extends to higher orders in $\l$: a solution at order 
$\l^{2n}$, 
$\{F_{1,2n}\,,\ G_{1,2n-1} \}$, is determined up to a constant 
$\calc_{1,n}$, which is being uniquely fixed at order $\calo(\l^{2(n+1)})$. 
} at order $\calo(\l^4)$:
\begin{equation}
\calc_{1,1}=\frac{3163}{421120}\,.
\eqlabel{c11} 
\end{equation}
\nxt $j=1$ level,
\begin{equation}
\c_{0}^{(1)}=10\,,\qquad \c_1^{(1)}=-\frac{2075}{288}\,.
\eqlabel{cj1}
\end{equation}
\nxt $j=2$ level,
\begin{equation}
\c_{0}^{(2)}=14\,,\qquad \c_1^{(2)}=-\frac{60613}{5760}\,.
\eqlabel{cj2}
\end{equation}
\nxt $j=3$ level,
\begin{equation}
\c_{0}^{(3)}=18\,,\qquad \c_1^{(3)}=-\frac{62451}{4480}\,.
\eqlabel{cj3}
\end{equation}

\subsection{Linearized fluctuations about $j=0$ boson stars}

\begin{figure}[t]
\begin{center}
\psfrag{c2}{$\c^2$}
\psfrag{p0h}{$p_0^h=\l$}
  \includegraphics[width=4in]{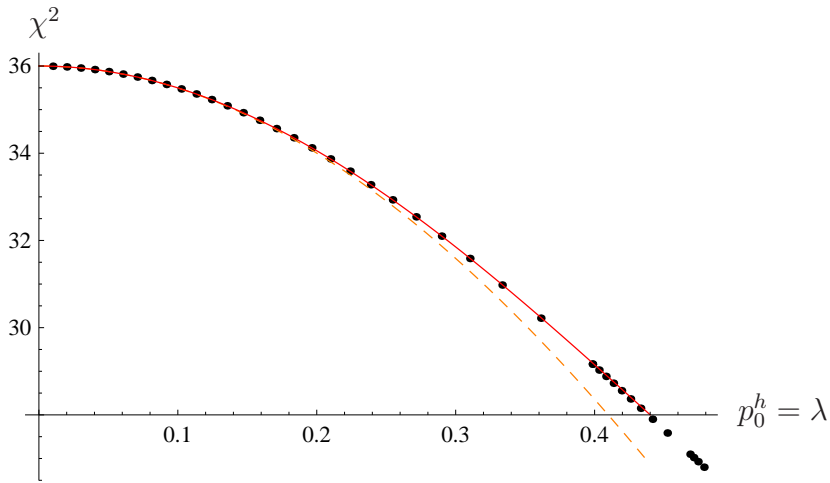}
\end{center}
  \caption{(Colour online) Spectrum of linearized fluctuations 
about $j=0$ boson stars as a function of $p_0^h$ (black dots).
The dashed orange/solid red curves are successive approximations to 
$\c^2=\left(\chi(p_0^h\equiv\lambda)\right)^2$ in $\l^2$, 
see \eqref{seriesg1f1}.     
}
\label{figurefl0}
\end{figure}

The spectrum of linearized fluctuations about $j=0$ boson stars 
is presented in Fig.~\ref{figurefl0}. We find that over the 
whole range of charges $Q$ we were able to construct $j=0$ 
boson stars, the frequency of their fluctuations squared ($\chi^2$) 
is positive. This strongly suggests that the ground state boson stars 
are perturbatively stable. 

As discussed in the previous section (see \eqref{cj1}-\eqref{cj3}), 
excited boson stars are perturbatively stable for small charge.
Our numerical simulations suggest that both 
the $j=0$ and the excited boson stars are nonlinearly stable.

\section{Nonlinear Results}\label{section4}
We take the constructed boson star solutions described above $\left\{ \phi(x),d(x),a(x),\omega \right\}$ and employ them to
provide initial data for our dynamical studies via 
\begin{eqnarray}
\phi_i & = & \frac{\phi} { \cos^2 x} \, \delta^1_i\,,\\
\Phi_i & = &  \frac{\phi'} { \cos x} \, \delta^1_i\,,\\
\Pi_i  & = & \frac{\omega \phi e^{d}}{a} \, \delta^2_i,
\label{eq:BSID}
\end{eqnarray}
where the metric functions are obtained by solving the constraints.
We confirm convergence of the obtained solutions (by monitoring the constraint residuals,
charge and mass conservation and self-convergence vs time)
as resolution is increased (see also ~\cite{Buchel:2012uh}).

\noindent{\bf \em Perturbed, Genuine Boson Stars:}~~~
We concentrate on studying the behavior of these solutions when perturbed, and
have considered various forms of perturbation with qualitatively similar results. 
For concreteness, we here present results obtained with
Gaussian perturbations 
parametrized as $G(x) = \epsilon e^{-\left(r-R_0\right)^2/\Delta^2}$  and add it to the
boson star solution  via
\begin{eqnarray}
\phi_i & = & \left[ {\phi}/ { \cos^2 x}  + G(x) \right] \delta^1_i\,,\\
\Phi_i & = & \left[ {\phi'}/{\cos x} + G'(x) \right]  \delta^1_i\,,\\
\Pi_i  & = & \left[ \omega \phi\left({ e^{d}}/{a}\right) +G' \right] \delta^2_i.
\end{eqnarray}
In analogy with previous studies, we set the amplitude of the Gaussian perturbation
to $\epsilon$. We note that because the constraints are solved numerically at the initial
time for $a(x,0)$ and $\delta(x,0)$, this perturbed initial data together with the obtained
metric variables is fully consistent with the equations of motion.
Because our numerical implementation is fully non-linear, it naturally probes
the nonlinear behavior of this system, and we are
particularly interested in the stability of these boson star solutions.

Recall that Ref.~\cite{Bizon:2011gg}
found that pulses of scalar field in AdS are unstable to black hole
formation, and subsequent arguments in~\cite{Dias:2011ss} supported this view of generic instability.
A reasonable expectation in light of those works is that
any perturbation of the boson star will behave in a similar way; that is,
one expects such a perturbation to travel back and forth between origin and
AdS boundary, sharpening with each pass,
leading eventually to BH formation. However, our early studies suggested quite
the opposite (for sufficiently small perturbations); that instead boson stars were stable~\cite{lieblingtalk}
prompting us to study this system more broadly and deeply.

For small perturbations, very long-lived, regular solutions were obtained describing
perturbed boson stars. For such
long-lived solutions, we monitored the metric functions (e.g. max$|1-A(x,t)|$ and $\delta(0,t)$)
and they showed no signs of instability to BH formation.
These observations prompted a thorough study of the instability in AdS; independent
work via perturbative studies also pointed out that AdS should be
stable for several families of solutions~\cite{Dias:2012tq}.

A variety of boson star solutions were studied, including members of levels 0, 1, and 3 (level 2 solutions
presented regularity issues near the AdS boundary and we defer such analysis for future work). All examples
appeared stable.
Interestingly, in
asymptotically flat scenarios, excited boson stars are generally unstable, 
radiating energy and settling into a ground state solution~\cite{2012LRR....15....6L}. 
In AdS, however, there is no way to rid itself of 
excess charge, which presumably explains their stability. Nevertheless, this property of AdS does not explain how 
the boson star can be immune to the weakly turbulent instability. Below, we present an argument to this end,
but first we discuss the behavior of a different family that lends support to our argument. 

\noindent{\bf \em Fake Boson Stars:}~~~
Of course numerical evolutions are limited to finite times, and so one cannot rule out
that instability will manifest after the code has been stopped or beyond the time
for which one trusts the results. To better assess the observed behavior, we compare these
long-lived solutions to a different family which can be considered 
``nearby'' in some sense. This family, which we refer to as 
{\em fake boson stars}, represents purely real initial data with the same mass and 
profile as their counterpart {\em genuine} boson stars. 
A fake counterpart of some boson star solution of \eqref{eq:BSID} is achieved via
the transformation
\begin{equation}
\phi_1^{\rm fake} = \phi_1^{\rm BS}, ~~~~~~
\Pi_1^{\rm fake} = \Pi_2^{\rm BS}, ~~~~~~
\Pi_2^{\rm fake} = 0 \,.
\label{eq:fake}
\end{equation}

\begin{figure}
\centerline{\includegraphics[width=5in]{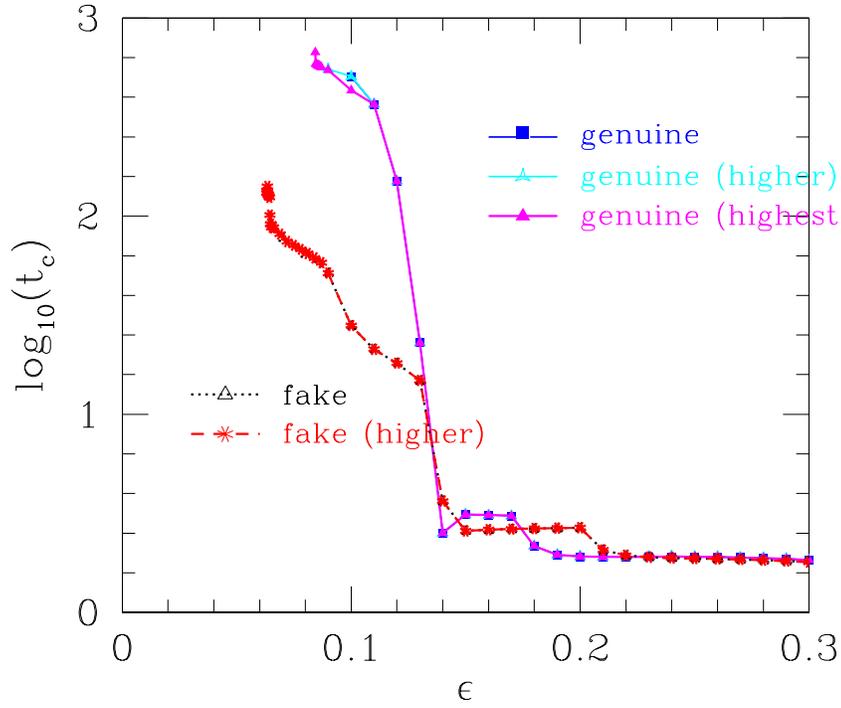}}
\caption{ (Colour online) Collapse times for Gaussian perturbations of a ground state boson
star ($\phi_1(0,0)=0.253$)
 and its corresponding fake star. Increasing resolutions are shown. For short collapse times, resolutions agree.
However, for the longest evolutions, higher resolutions are needed. Even with very high resolutions, small $\epsilon$
evolutions show no sign of collapse.
}\label{fig:gs31}
\end{figure}

\begin{figure}
\centerline{\includegraphics[width=5in]{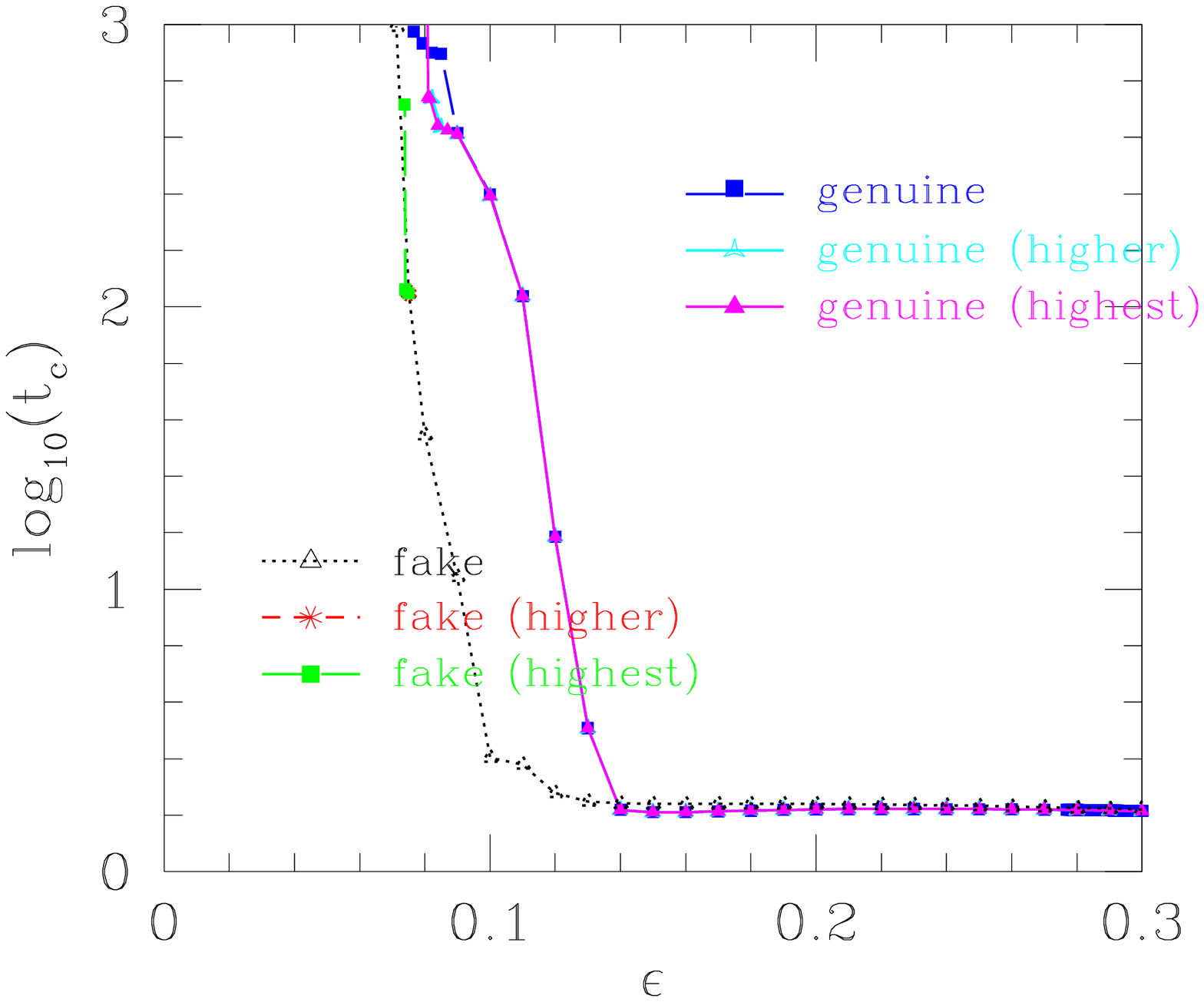}}
\caption{(Colour online) Collapse times for Gaussian perturbations of a first excited state boson star~($\phi_1(0,0)=-0.272$)
 and its corresponding fake star.
As in Fig.~\ref{fig:gs31}, higher resolutions are also shown with differences among the resolution appearing only at very late times.
}\label{fig:ex53}
\end{figure}

Remarkably, the evolution of this family also yields regular, long-lived solutions for small
perturbations that 
do not collapse to a black hole. Figs.~\ref{fig:gs31} and~\ref{fig:ex53} illustrate the time of collapse
as a function of  $\epsilon$ both for genuine and fake boson stars.
As indicated in the figures, successively higher resolutions largely coincide with differences only apparent at the latest times.
In all cases, the results indicate collapse times increasing quickly as $\epsilon$ decreases with no signs of collapse 
for smaller amplitudes of perturbation.

Notice that these fake solutions are not stationary and have no charge, two seemingly 
essential features of genuine boson stars, and so their apparent immunity to this weakly turbulent instability is surprising.
This ``stability'' is apparently not tied to special features (e.g. charge or stationarity) but instead
suggests that the dynamics undergoes something akin to a {\em frustrated resonance} in which amplitudes increase at times but then disperse. 
In particular, one essential aspect common to both genuine and fake boson stars appears to be their non-compact, long-wavelength nature. 
Because they have energy distributed throughout the domain, modes no longer propagate coherently. Instead there is a 
continuing competition between dispersion and gravitational contraction; collapse to a black hole or not is then
determined by the outcome of this competition.

\noindent{\bf \em Large $\sigma$:}~~~
Admittedly, this argument is far from rigorous. But if it holds, then
it would imply many other forms of stable initial data. In particular, 
perhaps other forms of initial data may be immune to this weakly-turbulent instability when its 
extent is large.
To explore this conjecture, we adopt the same form of data considered in many previous studies of this
instability~(such as those in~\cite{Bizon:2011gg,Buchel:2012uh,Jalmuzna:2011qw,Maliborski:2012gx,Liebling:2012gv}). 
We thus consider this family again, which takes the following form in our rescaled variables
\begin{equation}
\Phi_i(0,x)=0\,,\qquad \Pi_i(0,x)=\frac{2\epsilon}{\pi}
e^{-\frac{4\tan^2x}{\pi^2\sigma^2}}\ \cos^{1-d}x\  \delta_i^1\,.
\eqlabel{phiQz}
\end{equation}

\begin{figure}
\centerline{\includegraphics[width=5in]{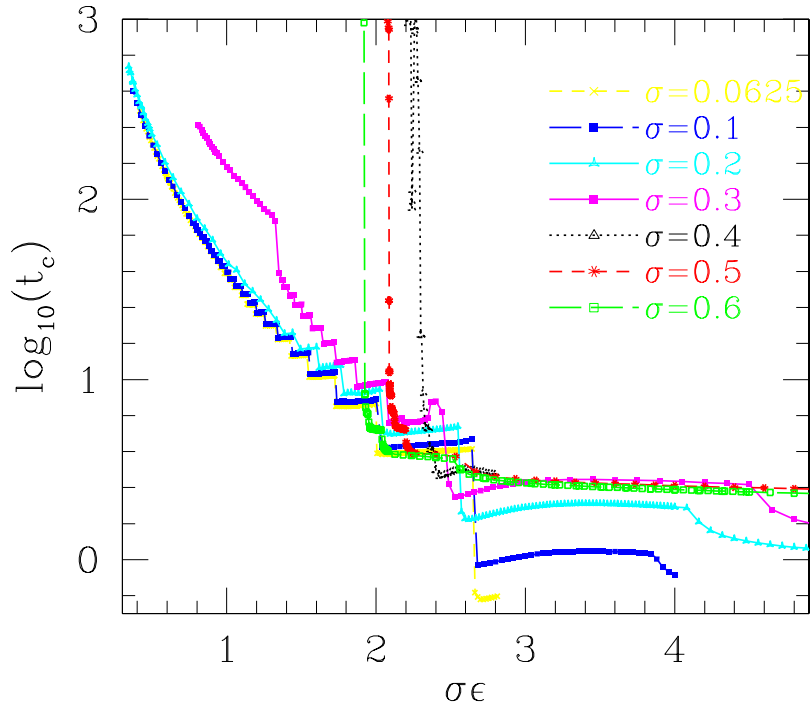}}
\caption{(Colour online) Collapse times for initial data of the form Eq.~\eqref{phiQz} with varying width values, $\sigma$.
Because changes to $\sigma$ affect the amount of mass, the natural parameter
against which to plot is $\sigma \epsilon$ not just $\epsilon$ (also see Fig.~\ref{fig:varyingsigma2} for this data plotted versus $\epsilon$). Note that
for $\sigma \lesssim 0.3$ the standard behavior is observed where collapse
eventually occurs for any $\epsilon$. In contrast for $\sigma \gtrsim 0.3$,
there appears to exist a threshold $\epsilon^*$ below which collapse does not occur.
For initial data above the transition, $\sigma >0.3$, evolutions with smaller $\epsilon$ than shown reached at least $t \approx 2000$ with no signs of eventually collapse.
}\label{fig:varyingsigma}
\end{figure}

\begin{figure}
\centerline{\includegraphics[width=5in]{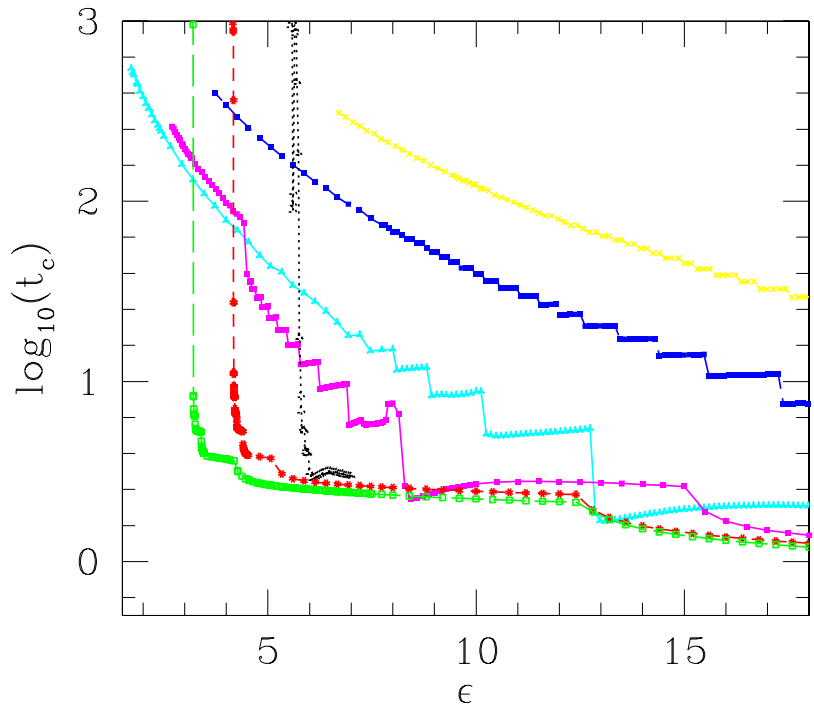}}
\caption{(Colour online) Same data as Fig.~\ref{fig:varyingsigma} but with an un-rescaled abscissa. As one's eye moves from upper-right down to lower-left, the initial data parameter $\sigma$ increases and the behavior of the time of collapse, $t_c$, changes dramatically from the ``usual'' stair-step to something else entirely with a very sharp transition. Note that the spacing between runs is very non-uniform and that tuning is necessary to resolve the transition.
}\label{fig:varyingsigma2}
\end{figure}

To test the possibility of regular development of this data, we 
considered the time development of Eq.~\eqref{phiQz} with varying values of
$\sigma$.  The results are plotted in Figs.~\ref{fig:varyingsigma} and~\ref{fig:varyingsigma2} which show
the time of collapse as a function of $\epsilon$ for various values of $\sigma$. As is evident from the figures, for
small values of $\sigma$ ($\sigma < 0.3$) the collapse time increases monotonically as $\epsilon$ decreases; however
for larger values of $\sigma$ the collapse time increases abruptly as $\epsilon$ is decreased.
Notice that this abrupt growth in collapse time behaves quite similarly to that 
seen for boson stars and fake stars, suggesting that for sufficiently large $\sigma$, the behavior would be regular.
Furthermore, an analysis of the Fourier power spectra of cases below $\sigma \approx 0.3$ reveal 
that the spectra monotonically shift to higher frequencies as time progresses. In contrast, for
cases above $\sigma \approx 0.4$ they do not do so. Instead the shift saturates and the spectral
content oscillates within a narrow window of frequencies.

That large-$\sigma$ initial data is immune to the weakly turbulent instability is consistent
with the argument that widely distributed mass energy prevents the coherent amplification
typical of the instability. It is interesting to consider what would happen in the
semilinear wave equation on a fixed AdS background as studied in~\cite{Liebling:2012gv}.
That model shows many of the same characteristics as the gravitating scalar collapse, but
the nonlinear potential plays the role of the attractive, focusing effect that gravity 
plays here. However, numerical evidence from that model suggests that there is no large-$\sigma$ effect, lending support to the idea that the distributed mass-energy affects the spacetime in a way that disturbs the coherent amplification.

\begin{figure}
\centerline{\includegraphics[width=5in]{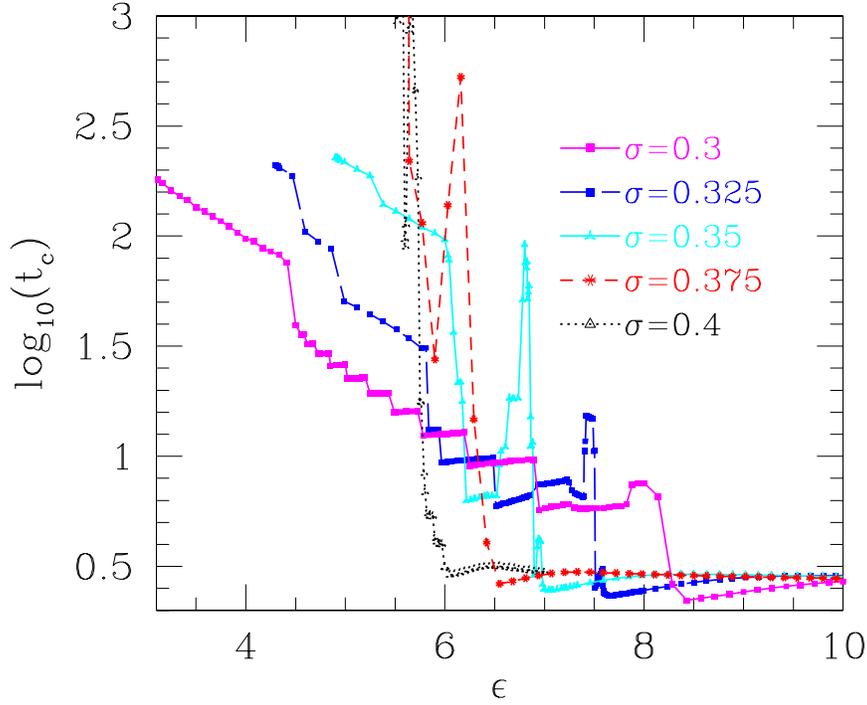}}
\caption{(Colour online) Collapse times for a range of $\sigma$ that demonstrates the transition from turbulent BH formation
 to frustrated resonance. For 
$\sigma=0.3$, collapse appears
inevitable for any value of $\epsilon$ in contrast to the results for $\sigma=0.4$. 
Interestingly, a ``bump'' appears for these values of $\sigma$ in which the collapse times demonstrate
a lack of monotonicity. Beginning with $\sigma=0.3$ (magenta, solid squares) around $\epsilon \approx 8$,
one sees a small bump that, as one looks to higher $\sigma$, sharpens and occurs at smaller $\epsilon$ values.
}\label{fig:sigmatransition}
\end{figure}

A change in behavior such as this merits a closer examination of the ``transition region'' between apparent
stability and black hole collapse. 
Fig.~\ref{fig:sigmatransition} illustrates this region $0.3\le \sigma \le 0.4$
in more detail. Interestingly, in this transition region
the time-to-collapse exhibits a seemingly oscillatory behavior
prior to displaying the characteristic rapid growth as $\epsilon$ is decreased.

From Figs.~\ref{fig:varyingsigma} and~\ref{fig:sigmatransition}, it is clear that for
$\sigma \ge 0.4$ there is some $\epsilon_{\rm min}$ below which initial data does
not form a black hole. The idea of this function $\epsilon_{\rm min}(\sigma)$
is similar to $\epsilon_{\rm min}(x_{\rm max})$ studied in~\cite{Liebling:2012gv}.
Preliminary study of $\epsilon_{\rm min}(\sigma)$ shows it to be a roughly exponentially
decreasing  function (after the apparent discontinuity at $\sigma \approx 0.4$ in which
$\epsilon_{\rm min} = 0 \rightarrow ~5.6$).
The behavior below $\epsilon_{\rm min}$ is demonstrated in Fig.~\ref{fig:dispersion}. In particular,
for this weak initial data when the metric is frozen at its initial profile, the evolution
demonstrates dispersion.

\begin{figure}[t]
\begin{center}
\psfrag{i}{$i$}
\psfrag{lc}{$\ln (c_i(\sigma))$}
  \includegraphics[width=4in]{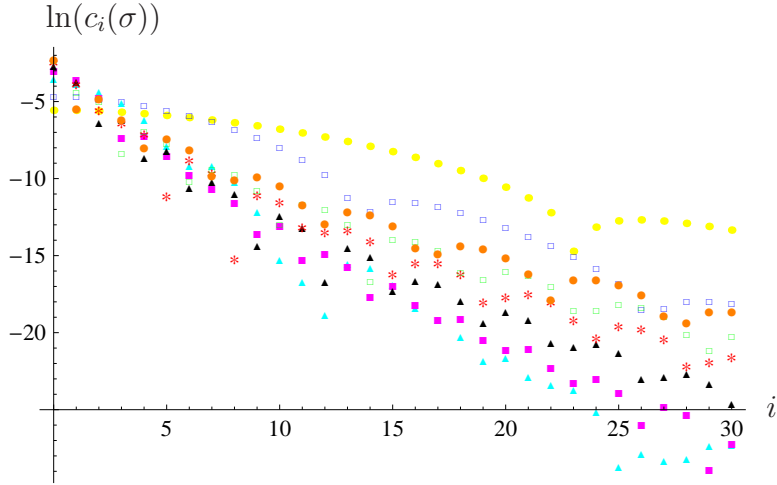}
\end{center}
  \caption{(Colour online) Spectral decomposition 
of the initial data \eqref{phiQz} with 
$\sigma=\{0.0625,0.1,0.2,0.3,0.4,0.5,0.6,0.7\}$ (yellow circles,
blue squares, cyan triangles, magenta squares, black triangles,
red stars, green squares, orange circles) and 
$\sigma\epsilon=1$ in the oscillon basis. Spectral coefficients $c_i(\sigma)$ 
(see \eqref{bizonspectral}) start relatively large  at high oscillon numbers for  
$\sigma=\frac{1}{16}$ (circles); they  decrease for $\sigma=0.1$ (squares), 
and achieve a
minimum profile around  $\sigma=0.2$ (triangles) or $\sigma=0.3$ (squares);
then they increase for $\sigma=0.4$ (triangles),
$\sigma=0.5$ (stars), $\sigma=0.6$ (squares) and $\sigma=0.7$  (circles). 
The minimum of the spectral coefficients profile roughly coincides with the 
critical value of $\sigma\sim 0.2-0.4$ which separates stable and unstable 
regions in the parameter space of the initial data   \eqref{phiQz}.
}
\label{figure9a}
\end{figure}

\begin{figure}
\centerline{\includegraphics[width=5in]{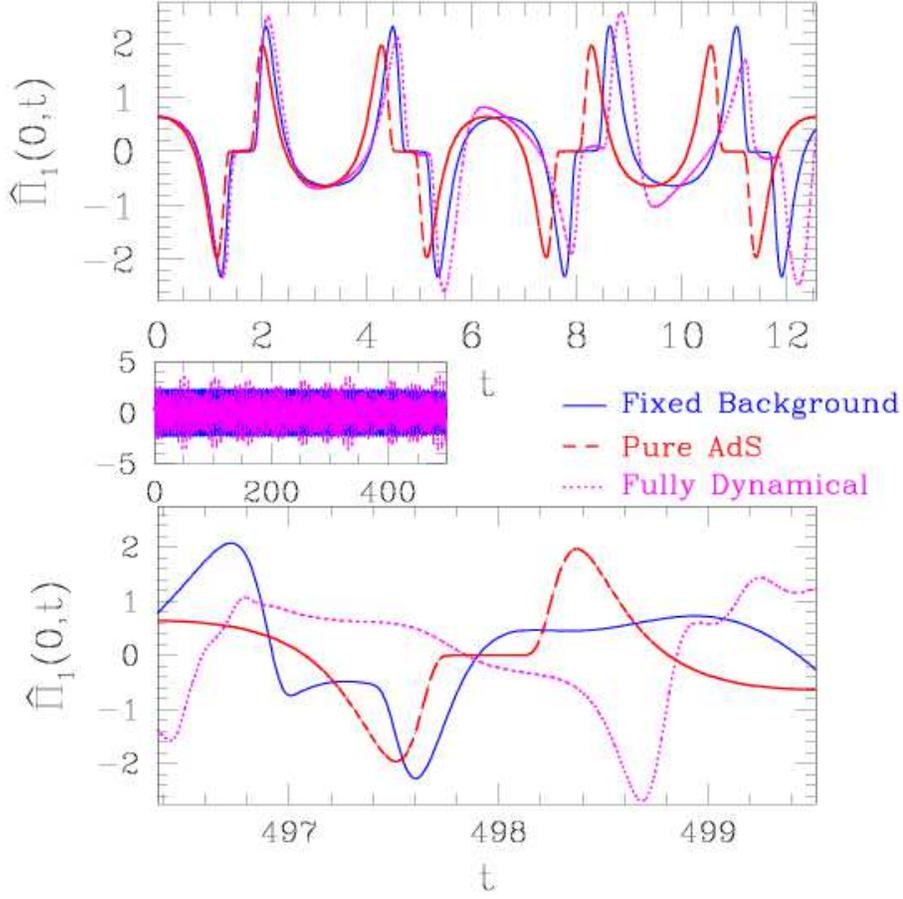}}
\caption{(Colour online) Demonstration of dispersion introduced by widely distributed mass-energy.  Shown is the behavior of $\Pi$ at the origin during the evolution of $\epsilon=1$, $\sigma=1$ initial data. By keeping $A=1$ and $\delta=0$, we evolve the
scalar field in a pure AdS background, and this results in a periodic solution (dashed, red line). Instead, by solving for the initial metric $A(x,0)$ and $\delta(x,0)$, we evolve the scalar field on a fixed background (solid, blue line).
This fixed-background evolution displays the dispersion introduced by the
fixed metric. The fully dynamical evolution (dotted magenta) is
also shown. In the intervening time (see inset), this solution shows the periods
of focussing and dispersal typical of what we call frustrated resonance.
Finally, at late times, we copy the pure AdS solution from the
period $0<t<\pi$ and display it shifted in time (dot-dashed black); that it overlays the 
pure AdS solution shows that scalar solutions on a background of AdS are
periodic.
This figure is similar to Fig.~5 of~\cite{Liebling:2012gv} which shows the dispersion introduced by a restricted domain.
}\label{fig:dispersion}
\end{figure}

It is instructive to study the spectral decomposition of the initial 
data \eqref{phiQz} in the oscillon basis for different $\sigma$. To relate with  the
analysis in Fig.~\ref{fig:varyingsigma}, we keep  $\sigma \epsilon=1$ fixed.
For a select set, \ie,  $\sigma=\{0.0625,0.1,0.2,\cdots 0.7\}$, we compute the
spectral coefficients $c_i(\sigma)$, see (\ref{pihdef}), 
\begin{equation}
c_i=\bigg|\frac{1}{\w^{(i)}}\int_0^{\pi/2} dx\ \tan^2 x\ A(0,x) e^{-\delta(0,x)} \Pi_1(0,x)\ e_i(x) \bigg|\,,
\eqlabel{bizonspectral}
\end{equation}
where $A(0,x)$ and $\delta(0,x)$ are obtained from integrating \eqref{constx} with 
initial data \eqref{phiQz}. The resulting spectral decompositions are collected  in Fig.~\ref{figure9a}.
Comparing with the spectral decomposition of boson stars (see Fig.~\ref{figure3a}),
here, the large-$j$ decay of the spectra is approximately exponential, 
instead of a power-law as in  \eqref{os3}. The spectral profile achieves a minimum around 
$\sigma\sim 0.2-0.4$, which is roughly the critical value of $\sigma$ separating the 
stable and unstable 
regions in the parameter space of the initial data \eqref{phiQz}.

\noindent{\bf \em Restricted domain:}~~~
The last nonlinear effect presented in this section concerns evolutions conducted
in a restricted domain. In particular, an artificial, reflecting  boundary condition
is applied at some $x_{\rm max} < \pi/2$, restricting the propagating pulse to
some subdomain of AdS. The motivation for this is to study whether this turbulent
instability is itself just the manifestation of the nonlinear attraction of gravity
occurring in a bounded domain, or instead some particular property of the full
AdS (and hence would be destroyed by this restriction).

As found in~\cite{Buchel:2012uh}, the imposition of such a reflecting boundary
condition does not eliminate black hole formation after multiple bounces. However,
a minimum value of $\epsilon$ was found, below which no such black hole formation
occurred. In the semilinear model, it was found that the boundary condition resulted
in dispersion not seen with the full AdS domain.

\begin{figure}
\centerline{\includegraphics[width=5in]{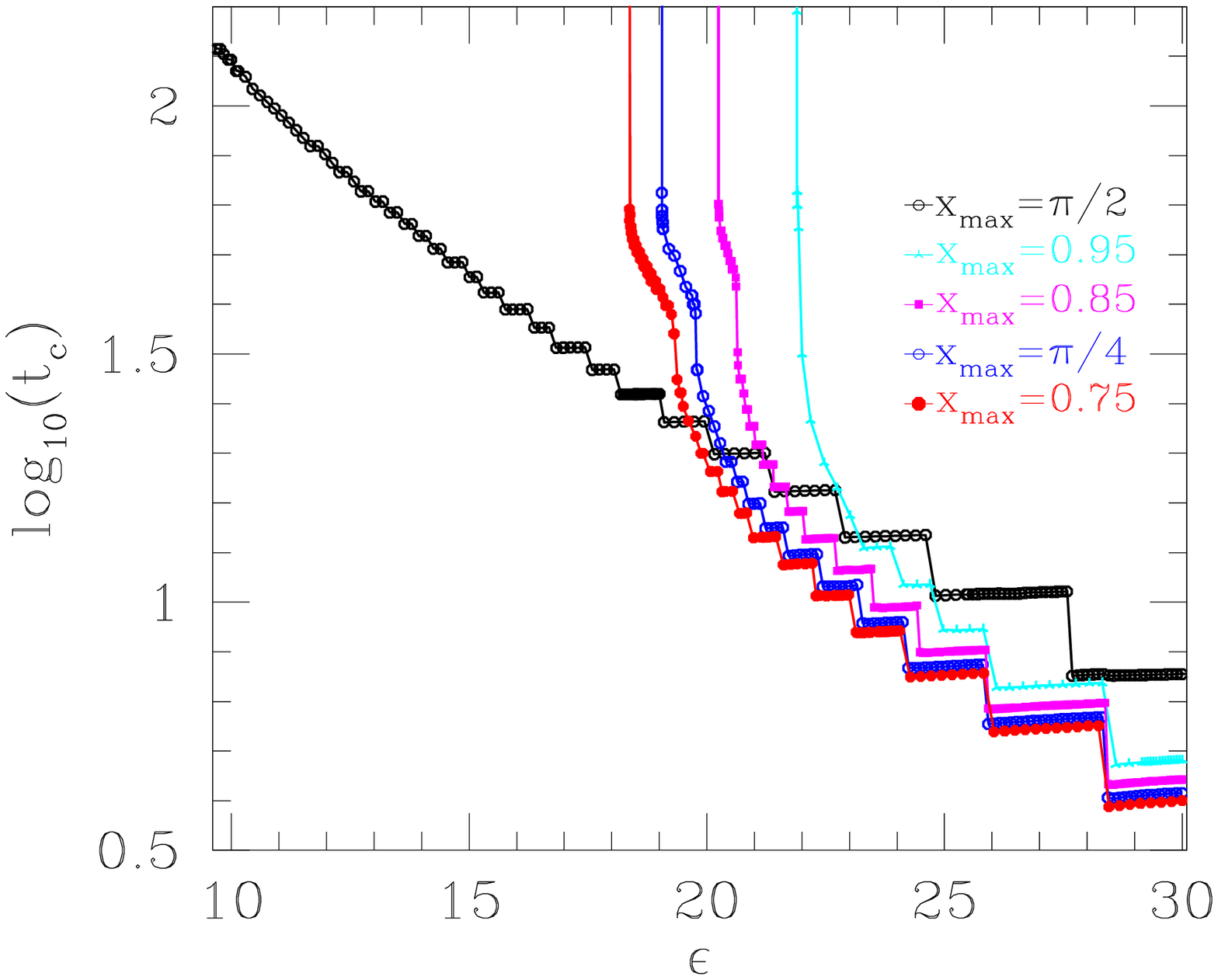}}
\caption{(Colour online) Collapse times of the initial data in Eq.~\eqref{phiQz} with $\sigma=1/16$ with an artificial, reflecting
wall at a various positions, $x_{\rm max}$. Also shown (solid black) are the 
results for the full domain with no reflecting wall. Note that the
finite domain evolutions demonstrate a threshold $\epsilon_{\rm min}$ below which collapse does not occur.
}\label{fig:finitedomain}
\end{figure}

Here, we revisit this problem, showing the customary time of collapse plot in
Fig.~\ref{fig:finitedomain}. As mentioned, there is some similarity with
the large-$\sigma$ effect in the existence of some $\epsilon_{\rm min}$. However,
in contrast, one does observe the ``stair-step'' decrease in collapse time for
$\epsilon$ increasing above $\epsilon_{\rm min}$, characteristic of successive
bounces.

It is also possible that such restricted-domain evolutions result from two effects:
(i) the imposition of the reflecting wall introduces dispersion as in the semilinear
model, and (ii) as the domain shrinks, there may be some effect due to the fact that,
for fixed $\sigma$, the fractional support of the initial data is increasing.

\section{Conclusions}\label{section5}
We have constructed boson star solutions in global AdS and shown
that they are stable at linear order. 
Numerical studies of their dynamics strongly suggest that these
boson stars, both ground state and the first few excited states, are non-linearly stable. Along with
solutions presented in~\cite{Maliborski:2013jca}, there is
now considerable evidence that the instability of AdS to scalar perturbations reported in~\cite{Bizon:2011gg}
is limited in scope; that there exist
non-trivial, dynamical examples of stable solutions
in AdS. 
These results are consistent with the perturbative arguments 
of~\cite{Dias:2012tq} for the stability of boson stars, geons
and solitons.

Comparison of the lifetimes of perturbed boson stars with other, non-stationary solutions, our
fake boson stars,  reveals an even wider class of initial data which appears to frustrate
the resonance and thereby avoid collapse for sufficiently small amplitude.
Indeed, using the very same family of the seminal work of~\cite{Bizon:2011gg},
numerical evolutions suggest that for initial data with $\sigma \gtrapprox 0.4$ the
instability can be avoided. More generally, initial data with widely distributed mass-energy
appears to be similarly immune to the turbulent instability.
This behavior is in contrast to what is
observed in solutions to the semilinear wave equation in (flat) AdS. There, a 
singularity always  forms~\cite{Liebling:2012gv}, even for large-$\sigma$, indicating that gravity 
plays a key role in the dynamics. In particular, a heuristic argument
suggests that the widely distributed mass-energy distorts the space
sufficiently to introduce dispersion and thereby oppose the concentrating
effect of the instability.

The picture that emerges is a phase space with (at least) two regions.
One region is subject to the weakly turbulent instability and therefore
collapses for any initial ``amplitude.'' The other region
can avoid the turbulent instability and
contains oscillons, boson stars, geons, and similar symmetric solutions.
However, it appears this second ``stable'' region contains a wider
class of solutions with no periodicity or stationarity, namely fake
boson stars, large-$\sigma$ and similarly distributed families of
initial data. To be clear, this second region need not be strictly stable
and can certainly possess collapsing solutions. What is important is that
in this second region, one can choose a sufficiently small amplitude such 
that the weakly turbulent instability is avoided. 

We do not know at this stage what is the precise criterion that separates 
the parameter region of scalar field initial data
resulting in BH formation from the region of non-linear stability. It can be argued 
that such a criterion is encoded in the spectral decomposition of the 
initial data in the oscillon basis. Indeed, consider the region
(assuming it exists\footnote{Numerical analysis of  
\cite{Bizon:2011gg} strongly supports that this is the case.}) where collapse occurs 
for arbitrarily small amplitude of the scalar field. In this limit, the full initial 
data is the oscillon spectrum, as the backreaction can be safely ignored. In this paper 
we presented  strong evidence for initial configurations that do not collapse 
in the limit of vanishingly small amplitude. Thus, the distinction between stable 
and non-stable configurations (at least for small amplitudes) must be hidden in 
the initial scalar field spectral data. We have seen that initial profiles 
of~\cite{Bizon:2011gg}  \eqref{phiQz}  become stable for small $\epsilon$ as 
$\sigma$ increases; the latter increase results in softening the decay of the 
asymptotic oscillon spectral coefficients (see Fig.~\ref{figure9a}). Likewise, boson stars have 
an asymptotic power-law oscillon spectral decomposition, in contrast to the exponential-decay profile 
for initial data \eqref{phiQz} (see Fig.~\ref{figure3a}). 

It is important to further investigate the nonlinear stability of AdS. The issue 
has profound implications for a dual boundary conformal field theory, as it identifies 
CFT initial configurations that fail to thermalize. There is no obvious symmetry criterion
``protecting'' such configurations. A possible future direction is to investigate the collapse of 
initial configurations specified by  their oscillon spectral decompositions with various trial 
profiles $c_i$. The primary goal, of course, is the identification of the stability criteria ---
it is possible that the latter can be established from analysis of the weakly nonlinear regime only.
Finally, it is interesting to analyze the stability of more general configurations ---
such as boson stars with  {\it local} bulk charge, as recently discussed in \cite{Gentle:2011kv,Hu:2012dx,Brihaye:2013hx}.

%
%
~\\
\noindent{\bf{\em Acknowledgments:}}
It is a pleasure to thank Oscar Dias, Chad Hanna, Gary Horowitz, Pavel Kovtun, Robert Myers, Andrzej Rostworowski and Jorge Santos
for interesting and helpful discussions.
This work was supported by the NSF (PHY-0969827 to Long Island University)
and NSERC through Discovery Grants (to AB and LL). LL and SLL thank the KITP 
for hospitality where parts of this work were completed.
Research at Perimeter
Institute is supported through Industry Canada and by the Province of Ontario
through the Ministry of Research \& Innovation.
Computations were performed thanks to allocations at the Extreme Science and Engineering Discovery Environment (XSEDE),
which is supported by National Science Foundation grant number OCI-1053575 as well as SHARCNET.

\bibliographystyle{utphys}     
\bibliography{./bs}

\end{document}